\documentclass[aps,prb,amsmath,amssyb,floatfix,twocolumn]{revtex4}
\usepackage{graphicx}% Include figure files
\usepackage{hyperref}
\usepackage{braket}
\usepackage{bm}

\def\LSCO{La$_{2-x}$Sr$_x$CuO$_4$ }
\def\BSCCO{Bi$_2$Sr$_2$CaCu$_2$O$_{8+\delta}$}
\def\YBCO{YBa$_2$Cu$_3$O$_{6+x}$}
\def\LBCO{La$_{2-x}$Ba$_{x}$CuO$_4$}
\def\prb{Phys. Rev. B }
\def\prl{Phys. Rev. Lett. }

\def\ie{{\it i.e.}}

\def\etal{{\it et al.}}
\def\bnabla{\bm{\nabla}}
\def\rhos{\rho_s}
\def\D{\mathcal{D}}
\def\na{{n_\alpha}}

\def\br{{\bf r}}
\def\bp{{\bf p}}

\def\dr{{d^2r}}
\def\Tr{{\rm Tr}}
\def\psisq{|\psi|^2}
\def\ec{\varepsilon_c}

\begin{document}

\title{Signatures of thermally excited vortices in a superconductor
  with competing orders}

%%\author{Gideon~Wachtel and Dror Orgad}
\author{Gideon~Wachtel}
\author{Dror~Orgad}

\affiliation{Racah Institute of Physics, The Hebrew University,
  Jerusalem 91904, Israel}

\date{\today}

\begin{abstract}
  Experimental evidence for the existence of a fluctuating
  charge-density wave order in the pseudogap regime of \YBCO~has
  renewed interest in its interplay with superconductivity. Here, we
  consider the problem within a nonlinear sigma model, which was
  recently proposed to describe the apparent competition between the
  two order parameters. In particular, we use a saddle-point
  approximation to calculate the properties of superconducting vortex
  excitations within such a model. In addition, we analytically
  calculate a collection of experimentally observable quantities,
  which probe both the superconducting and charge-density wave
  fluctuations, and identify expected signatures of thermally excited
  vortices.
\end{abstract}


\maketitle

\section{Introduction}

Over the past two decades, a large number of experiments have produced
evidence that the pseudogap state of the cuprate high-temperature
superconductors exhibits fluctuations towards various types of
order.\cite{intertwined} In particular, recent X-ray scattering
experiments\cite{Ghiringhelli,Chang,Achkar1,Blackburn13,Achkar2,
  Comin1,daSilva,Le-Tacon,Comin2,Hucker,Blanco-Canosa} indicate that
the pseudogap regime of underdoped \YBCO~is characterized by a local
charge-density wave (CDW) order. The situation concerning the nature
of superconducting (SC) fluctuations in the same compound is less
clear. On the one hand, $c$-axis infrared spectroscopy\cite{Dubroka}
has found signatures of a precursor SC state, which onsets close in
temperature to the CDW signal. In addition, high-field torque
magnetometry\cite{Yu2014} has detected fluctuation diamagnetism up to
high temperatures. On the other hand, a different magnetometry
measurement\cite{Kokanovic} found diamagnetic response, consistent
with Gaussian SC fluctuations, only in a narrow range above the
critical temperature, $T_c$. The same conclusion was reached based on
a measurement of the Nernst effect.\cite{Taillefer-broken-symm-Nature}

In any event, all available data point to the fact that the strength
of the CDW fluctuations is anti-correlated with
superconductivity. Specifically, the intensity of the CDW scattering
peak grows as the system is cooled towards $T_c$, and then decreases
upon entering the SC phase. Furthermore, the CDW signal is enhanced
when a magnetic field is used to quench superconductivity. Finally,
optical excitation of apical oxygen vibrations promotes transient
superconducting signatures,\cite{Hu,Kaiser} resembling similar results
in \LBCO, where they were attributed to the melting of stripe order.
Motivated by these findings Hayward \etal\cite{Hayward1,Hayward2} have
recently proposed a phenomenological non-linear sigma model (NLSM),
which formulates the competition between fluctuating SC and CDW order
parameters.  Similar models emerge also from more microscopic
considerations.\cite{Metlitski,Efetov,Meier} Using Monte-Carlo
simulations of their model, Hayward \etal~calculated the temperature
dependence of the X-ray structure factor and of the diamagnetic
susceptibility, and compared them to data from experiments on \YBCO.

The purpose of the present paper is to analytically study the
properties and expected experimental signatures of thermally excited
SC vortices, within the NLSM of Ref. \onlinecite{Hayward1}. In
conventional BCS superconductors the core of the vortex consists of
the normal metallic state. Consequently, the energy needed to create
the core is of the order of the Fermi energy, thus making thermally
excited vortices highly unlikely. However, this need not be the case
when superconductivity competes with another state of comparable
energy, as assumed in the NLSM. Such "cheap" vortices are required if
one attempts to explain the diamagnetic and Nernst signals of
underdoped cuprates within a vortex
picture.\cite{Lee,Nernst-vortex}. Experimentally, the checkerboard
halos observed around vortices in a magnetic field\cite{Hoffman} give
evidence that the vortex core may actually harbor local CDW order.

To make progress towards our goal we consider the NLSM in the limit of
a large number, $N$, of fluctuating fields, and construct an effective
theory for the SC vortices by integrating out the CDW degrees of
freedom. Using the resulting theory we estimate the vortex core radius
and core energy, from which we determine the density of thermally
excited vortices. This allows us to identify the temperature range
above the transition temperature over which vortices remain well
defined and the physics is dominated by SC phase fluctuations. In this
temperature range we calculate the magnetization, $M_z$, and the
transverse thermoelectric transport coefficient, $\alpha_{yx}$, which
is related to the Nernst signal.\cite{Onglongprb} We find that both
decay rapidly with temperature in a manner that is governed by the
vortex core energy, while, on the other hand, the proliferation of
vortices leads to a rise of the X-ray structure factor, $S_{CDW}$.  As
the temperature is increased further the system crosses over to a
regime where the vortices are no longer well defined, amplitude
fluctuations become important, and the fluctuations are nearly
Gaussian. In this regime $M_z$ and $\alpha_{yx}$ continue to decrease,
albeit in a more moderate fashion, and $S_{CDW}$ also becomes a
decreasing function of temperature, thereby implying the existence of
a peak.

The paper is organized as follows. In section \ref{sec:model} we
present the model, consider its large-$N$ limit and identify
the various temperature regimes which emerge. The derivation of
an effective theory for the SC field is presented in section \ref{sec:core},
which also contains a calculation of the vortex core energy and core size,
as well as a numerical solution of the vortex structure. Section
\ref{sec:nernst} discusses the expected diamagnetic and Nernst signals
in different temperature regimes, while section \ref{sec:xray}
describes the maximum in the X-ray structure factor as a
function of temperature. We conclude with a discussion of the 
relation to experiments in section \ref{sec:conc}.

\section{The model and its large-$N$ effective theory}
\label{sec:model}

Hayward \etal\cite{Hayward1} considered a real 6-dimensional order
parameter, equivalent to a complex SC field $\Psi=n_1+in_2$ and two
complex CDW fields, $\Phi_x=n_3+in_4$ and $\Phi_y=n_5+in_6$. In this
paper we would like to use a saddle point approximation for the CDW
fields, which is formally justified when their number is large.  Thus,
we analyze a system described by a complex SC field $\{\psi,\psi^*\}$,
and $N-2$ real CDW fields $\{\na\}$, where $\alpha=1\dots N-2$. For
the sake of simplicity we disregard quartic and anisotropic CDW terms,
which appear in the Hamiltonian of Ref. \onlinecite{Hayward1}, and
analyze the more basic form
\begin{equation}
  \label{eq:H}
  H = \frac{\rhos}{2}\int\dr\left\{|\bnabla\psi|^2
    + \sum_{\alpha=1}^{N-2}\left[\lambda(\bnabla\na)^2+g n_\alpha^2\right]\right\}.
\end{equation}
Here $\rhos$ is the stiffness of the SC order, $\lambda\rhos$ is the
corresponding quantity for the CDW components, and $g\rhos$ is the
energy density penalty for CDW ordering. Central to the model is the
assumption that some type of order (SC or CDW) is always locally
present, in the sense of its amplitude, but that the different order
parameters compete, as expressed by the constraint\cite{constraint}
\begin{equation}
  \label{eq:cons}
  |\psi|^2+\sum_{\alpha=1}^{N-2}n_\alpha^2=N.
\end{equation}

A free energy functional $F[\psi^*,\psi]$ for the SC field is obtained
by integrating out the CDW fields
\begin{eqnarray}
  \label{eq:F}
 \!\!\!\!\!\!\!\!\!\! e^{-\beta F} \!&=& \!\int\D\na\delta\left(|\psi|^2
  +\sum_{\alpha=1}^{N-2}n_\alpha^2-N\right)e^{-\beta H}\nonumber \\
  \!&=&\! \int\D\na\D\bar\sigma e^{-\beta H +i \int \dr \,\bar\sigma\left(|\psi|^2
      +\sum_\alpha n_\alpha^2-N\right)},
\end{eqnarray}
where $\beta=1/T$. In the limit $N\to\infty$ we carry out the
integration over $\na$ while assuming that the Lagrange multiplier
field $\bar\sigma$, which enforces the constraint, is fixed at its
saddle point configuration $\bar\sigma=-i\sigma$. As a result
\begin{eqnarray}
  \label{eq:Trln}
  \beta F & = & \frac{N-2}{2}\Tr\ln\left[\frac{1}{2}\beta\rhos\left(
    -\lambda\nabla^2+ g\right)+\sigma\right] \nonumber \\
  & & +   \int\dr\left[\frac{1}{2}\beta\rhos|\bnabla\psi|^2
  +\sigma\left(|\psi|^2-N\right)\right],
\end{eqnarray}
where $\sigma$ is determined by the saddle point equation
\begin{eqnarray}
  \label{eq:SP}
  \frac{\delta\,\beta F}{\delta\sigma(\br)} & = &
  \frac{N-2}{2}\Tr\left[\left(\frac{1}{2}\beta\rhos\left(
    -\lambda\nabla^2 +g\right)+\sigma\right)^{-1}
  \delta_\br\right]\nonumber \\ & &+\psisq-N=0,
\end{eqnarray}
with $\delta_\br$ an operator, whose functional form is
\begin{equation}
  \label{eq:delta}
  \delta_\br(\br',\br'')=\delta(\br'-\br)\delta(\br''-\br).
\end{equation}
The most likely SC configurations, those that minimize the free
energy, are determined by the second saddle point equation
\begin{equation}
  \label{eq:SPpsi}
  \frac{\delta\,\beta F}{\delta\psi^*} = -\frac{1}{2}\beta\rhos\nabla^2
  \psi+\sigma\psi=0.
\end{equation}
Below the mean-field transition temperature, $T_{MF}$,
Eqs. (\ref{eq:SP},\ref{eq:SPpsi}) acquire a uniform solution
$\psi(\br)=\psi_0$, $\sigma=0$ with
\begin{equation}
  \label{eq:psi2}
  |\psi_0|^2=N\left(1-\frac{T}{T_{MF}}\right),
\end{equation}
and
\begin{eqnarray}
  \label{eq:TMF}
  \!\!\!\!\!\!\!\!\!\frac{\rhos}{T_{MF}} & = & \frac{N-2}{N\lambda}{\rm Tr}\left[\left(
      -\nabla^2+g/\lambda\right)^{-1}\delta_{\br}\right] \nonumber \\
  & = & \frac{N-2}{N\lambda}\int\frac{d^2p}{(2\pi)^2}
  \frac{1}{p^2+g/\lambda} \nonumber \\
  & \simeq & \frac{N-2}{4\pi N\lambda}\left\{\begin{array}{ccc}
  \ln\left(32\lambda/ga^2\right) & & :\lambda/ga^2\gg 1 \\
  4\pi\lambda/ga^2+{\mathcal O}(\lambda^2) & & :\lambda/ga^2\ll 1
  \end{array}\right..
\end{eqnarray}
Here, and wherever is needed in the following, we regularize the
theory by putting it on a square lattice with lattice constant
$a$. This amounts to replacing the Laplacian by its discrete version
$p^2\rightarrow[4-2\cos(p_xa)-2\cos(p_ya)]/a^2$ and extending the
momentum integration over the first Brillouin zone $|p_{x,y}|<\pi/a$.

Beyond the mean-field approximation $T_{MF}$ is only a crossover
temperature, below which the most likely value of $\psi(\br)$ assumes
a finite amplitude.  However, phase fluctuations, particularly in the
form of vortices, prevent ordering down to a lower
Berezinskii-Kosterlitz-Thouless\cite{KT} temperature $T_{BKT}$.
$T_{BKT}$ itself can be estimated using Monte-Carlo results\cite{XYTC}
for the $XY$-model on a square lattice, which when applied to our
model gives $T_{BKT}\approx 0.9\rhos|\psi_0|^2$. Combined with
Eq. (\ref{eq:psi2}) this implies $T_{MF}/T_{BKT}=1+T_{MF}/0.9\rhos N$
and therefore the existence of a phase fluctuations regime, provided
that our results hold down to $N=6$.

\begin{figure}[t]
  \centering
  \includegraphics[width=\linewidth]{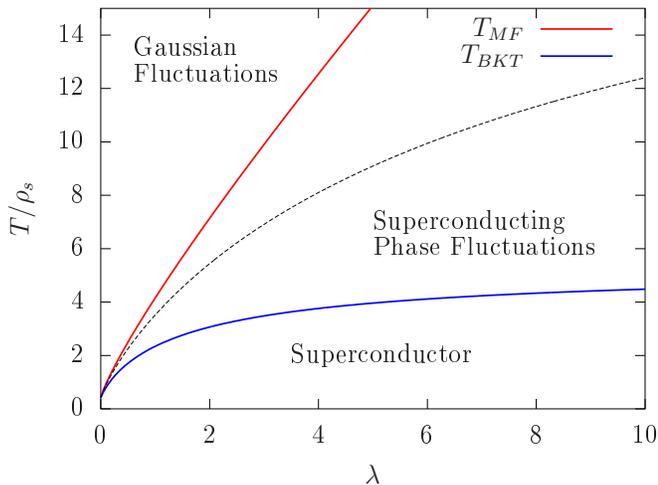}
  \caption{Phase diagram as a function of $\lambda$ and temperature,
    $T$, obtained by extending our large $N$ results down to $N=6$.
    Here, $ga^2=0.3$. At low temperatures, $T<T_{BKT}$, the system is
    superconducting, while Gaussian fluctuations approximately
    describe the high temperatures regime, $T>T_{MF}$. Above $T_{BKT}$
    the system enters a phase fluctuations regime, with well defined
    vortices up to an intermediate temperature, depicted by the dotted
    line. We define this curve as the temperature where
    $n_fr_0^2\approx 0.5$, and approximate it using
    Eqs. (\ref{eq:EcestlowT}) and (\ref{eq:nf}).}
  \label{fig:phasedia}
\end{figure}

Therefore, it is possible to construct a schematic phase diagram, Fig.
(\ref{fig:phasedia}), in which we identify three temperature regions:
(i) A high temperature regime, $T>T_{MF}$, approximately described by
Gaussian fluctuations in both the SC and CDW fields; (ii) a
superconducting phase fluctuations regime, $T_{BKT}<T<T_{MF}$, with
thermally excited vortices; and, (iii) a SC phase for $T<T_{BKT}$.
Eq. (\ref{eq:TMF}) indicates that for our case of interest,
$\lambda/ga^2>1$, $T_{MF}$ grows approximately linearly with $\lambda$
and depends only weakly on $g$. Thus, the size of the phase
fluctuations regime, (ii), is also expected to increase with
$\lambda$. We note, however, that as the temperature is increased in
this region, thermally excited vortices become denser and cease to be
distinct objects. Consequently, significant amplitude fluctuations,
associated with abundant vortex cores, are expected already at
temperatures below $T_{MF}$. A more stringent definition of the phase
fluctuations regime, would therefore require that the distance between
thermally excited vortices be strictly larger than their size, \ie,
$n_f< r_0^{-2}$. Even so, we still find that the extent of this regime
grows with $\lambda$, as depicted by the dotted line in Fig.
\ref{fig:phasedia}.

\section{The Vortex Core}
\label{sec:core}

\subsection{Effective Ginzburg-Landau theory and the vortex core size}

In the SC phase fluctuations regime, observables, such as the
magnetization, $M_z$, and transverse thermoelectric transport
coefficient, $\alpha_{yx}$, depend on the density, $n_f$, of thermally
excited vortices.\cite{Halperin,Benfatto-magnetic,
  Oganesyan,Nernst-vortex} This density, in turn, is set by the
energy, $\ec$, and the linear size, $r_0$, of the vortex
core.\cite{MinnhagenRMP} Thus, in order to calculate observables in
this temperature regime, one must first know the temperature
dependence of the vortex core energy and size.

In order to estimate the vortex core size it is useful to construct a
Ginzburg-Landau type theory for small values of $\psi$, which is
applicable near the transition temperature $T_{MF}$, or, as in our
case, at lower temperatures, but near the vortex center. When $\psi$
is small it is possible to write the free energy $F$ explicitly by
expanding Eq. (\ref{eq:Trln}), and simultaneously solving the saddle
point equation (\ref{eq:SP}). For convenience, we define $\Sigma_0$
and $\Sigma_1$ by
\begin{equation}
  \label{eq:def01}
  \frac{1}{2}\beta\rhos\lambda(\Sigma_0+\Sigma_1)=\frac{1}{2}
  \beta\rhos g + \sigma,
\end{equation}
where, $\Sigma_0\sim\mathcal{O}(1)$ and
$\Sigma_1\sim\mathcal{O}(\psisq/N)$. We also introduce the
operator $K$
\begin{equation}
  \label{eq:K}
  K=(-\nabla^2+\Sigma_0)^{-1},
\end{equation}
in terms of which the saddle point equation (\ref{eq:SP})
takes the form
\begin{equation}
  \label{eq:SP01}
  \frac{N-2}{\beta\rhos\lambda}\Tr\left[\left(K^{-1}+\Sigma_1\right)^{-1}
  \delta_\br\right]=N-\psisq.
\end{equation}
$\Sigma_0$ itself is set by the zeroth order expansion of this
equation in $\psisq/N$,
\begin{equation}
  \label{eq:SP0}
  \frac{N-2}{\beta\rhos\lambda}\Tr[K\delta_\br]=
  \frac{N-2}{\beta\rhos\lambda}\int\frac{d^2p}{(2\pi)^2}\frac{1}{p^2+\Sigma_0}=N.
\end{equation}
Regularizing the integral on the lattice we obtain for $T<T_{MF}$
\begin{equation}
   \label{eq:Sigma0}
   \Sigma_0=\frac{32}{a^2}
   \exp\left(-\frac{4\pi N\lambda}{N-2}\frac{\rhos}{T}\right).
\end{equation}
To first order in $\psisq/N$, Eq. (\ref{eq:SP01}) implies the relation
\begin{equation}
  \frac{N-2}{\beta\rhos\lambda}
  \Tr\left[K\Sigma_1K\delta_\br\right] = \psisq,
\end{equation}
whose Fourier transform reads
\begin{equation}
\label{eq:Sigma1Fourier}
  \Sigma_1({\bf q})\int\frac{d^2p}{(2\pi)^2}
  \frac{1}{p^2+\Sigma_0}\frac{1}{({\bf p} +{\bf q})^2+\Sigma_0}
  =\frac{\beta\rhos\lambda}{N-2}|\psi|^2({\bf q}),
\end{equation}
leading for small $q$ to
\begin{equation}
\label{eq:Sigma1sol}
  \Sigma_1=\frac{4\pi\beta\rhos\lambda}{N-2}
  \left(\Sigma_0-\frac{1}{6}\nabla^2\right)\psisq.
\end{equation}

In terms of $\Sigma_0$, $\Sigma_1$ and $K$, the free energy,
Eq. (\ref{eq:Trln}), is written as
\begin{widetext}
\begin{equation}
  \beta F = \frac{N-2}{2}\Tr\ln\left[\frac{1}{2}\beta\rhos\lambda
    (K^{-1}+\Sigma_1)\right]+\frac{1}{2}\beta\rhos\int\dr\Big[
    |\bnabla\psi|^2+\left(\lambda\Sigma_0+\lambda\Sigma_1
      -g\right)\left(\psisq-N\right)\Big].
\end{equation}
Expanding the trace in orders of $\Sigma_1$,
\begin{eqnarray}
  \Tr\ln\left[\frac{1}{2}\beta\rhos\lambda
    (K^{-1}+\Sigma_1)\right]
  & = & \Tr\ln\left[\frac{1}{2}\beta\rhos\lambda K^{-1}\right] +
  \Tr[K\Sigma_1]-\frac{1}{2}\Tr[K\Sigma_1K\Sigma_1]+\cdots,
\end{eqnarray}
and using Eqs. (\ref{eq:SP0}) and (\ref{eq:Sigma1sol}) we finally obtain
a Ginzburg-Landau type free energy for $\psi$,
\begin{eqnarray}
  \label{eq:GL}
  F & = & \frac{1}{2}\rhos\int\dr\left[|\bnabla\psi|^2
    +(\lambda\Sigma_0-g)\psisq+\frac{2\pi\beta\rhos\lambda^2}{N-2}
    \Sigma_0|\psi|^4\right],
\end{eqnarray}
\end{widetext}
where we have neglected gradients in the quartic term.

The theory described by Eq. (\ref{eq:GL}) has a mean-field transition
when $\lambda\Sigma_0-g=0$, which is consistent with
Eq. (\ref{eq:TMF}).  In addition, the linear size, $r_0$, of the
vortex core in such a theory is roughly given by
\begin{equation}
  \label{eq:r0}
  r_0^{-2}\sim-(\lambda\Sigma_0-g).
\end{equation}
According to Eq. (\ref{eq:Sigma0}), at low temperatures, $\Sigma_0$ is
exponentially small, and $r_0^{-2}\sim g$ is independent of
temperature.

\subsection{Condensation energy and vortex core energy}

The vortex core energy roughly scales as
\begin{equation}
  \label{eq:Ecest}
  \ec\sim U r_0^2,
\end{equation}
where $U$ is the condensation energy density, \ie, the difference in
free energy density between a state where SC is uniformly condensed,
$\psi(\br)=\psi_0$ and a state which is uniformly non-SC,
$\psi(\br)=0$.  From Eq. (\ref{eq:Trln}) it follows that for the SC
saddle point solution, $\psi(\br)=\psi_0$, $\sigma=0$, the free energy
density is
\begin{equation}
  \label{eq:F0}
  \frac{F[\psi(\br)=\psi_0]}{L^2} = \frac{(N-2)T}{2L^2} \Tr\ln
  \left[-\lambda\nabla^2+ g\right],
\end{equation}
with $L^2$ the system area.  When $\psi(\br)=0$, $\sigma$ assumes a
different value, $\sigma=\beta\rhos(\lambda\Sigma_0-g)/2$, with
$\Sigma_0$ given by Eq. (\ref{eq:Sigma0}). Substitution into
Eq. (\ref{eq:Trln}) leads to
\begin{eqnarray}
  \label{eq:F1}
  \frac{F[\psi(\br)=0]}{L^2} & = & \frac{(N-2)T}{2L^2} \Tr\ln
  \left[-\lambda\nabla^2 + \lambda\Sigma_0\right] \nonumber \\
  & &   -\frac{1}{2}\rhos(\lambda\Sigma_0-g) N.
\end{eqnarray}
Combining the two results we find that the condensation energy density
is given by
\begin{eqnarray}
  \label{eq:DeltaF}
  U & = & (F[\psi(\br)=0] - F[\psi(\br)=\psi_0])/L^2 \nonumber \\
  & = & \frac{(N-2)T}{2L^2}\,\Tr\left[\ln
  \left(-\lambda\nabla^2 + \lambda\Sigma_0\right)
  -\ln \left(-\lambda\nabla^2+g\right)\right] \nonumber \\
  && -\frac{1}{2}\rhos(\lambda\Sigma_0-g) N.
\end{eqnarray}

As noted above, $\Sigma_0\to 0$ at low temperatures. Therefore, in
this limit $U$ behaves according to
\begin{equation}
  \label{eq:dFlowT}
  U \approx  \frac{1}{2}\rhos g N-\frac{(N-2)T}{2}
  \int\frac{d^2p}{(2\pi)^2}\ln\left[\frac{p^2+g/\lambda}{p^2}\right].
\end{equation}
In the same temperature regime, one finds from Eq. ({\ref{eq:r0}) that
  $r_0^{-2}\approx g$.  Consequently, the main temperature dependence
  of $\ec$ originates from $U$, which decreases linearly with $T$
\begin{equation}
  \label{eq:EcestlowT}
  \frac{\ec}{\rhos}\sim\frac{N}{2}\left(1-b\frac{T}{\rhos}\right),
\end{equation}
where
\begin{equation}
  \label{eq:b}
  b=\frac{N-2}{2N}
  \int\frac{d^2p}{(2\pi)^2}\ln\left[\frac{p^2+g/\lambda}{p^2}\right]=
  \frac{\rhos}{T_{MF}}+\frac{N-2}{4\pi N\lambda}.
\end{equation}

\subsection{Numerical solution of the vortex structure}

Although the above analytical estimate yields the general behavior of
$\ec$ as a function of $T$, it cannot give $\ec$ in absolute values,
since we do not know the correct proportionality constant which enters
Eq. (\ref{eq:Ecest}). In order to bridge this gap and check the
validity range of the estimate, Eq. (\ref{eq:EcestlowT}), we have
calculated the vortex structure and energy numerically.  The vortex
configuration was obtained by solving the saddle point equations,
(\ref{eq:SP}) and (\ref{eq:SPpsi}), while imposing a phase winding
$\psi(\br)=f(r)e^{i\theta}$, where $\br$ is the position relative to
the vortex center, $r=|\br|$, and $\theta$ is its angle with respect
to the $x$ axis. For the solution we have used polar coordinates and
discretized the radial coordinate $r$ in units of a short distance
cutoff $a$. Figure \ref{fig:Nvortex}a shows the amplitude $f(r)$ as a
function of the distance from the vortex center, for a number of
temperatures below $T_{MF}$. In the calculation we have set $N=6$,
$ga^2=0.03$ and $\lambda=1$, which give $T_{MF}\approx 2.7\rhos$.

The vortex core energy was calculated by plugging the vortex solution
into Eq. (\ref{eq:Trln}), subtracting from it the free energy of the
uniform solution and the kinetic energy contribution $\rhos/2\int \dr
f(r)^2/r^2$ of the superflow around the vortex core. The squares in
Fig. \ref{fig:Nvortex}b depict the core energies for the vortex
structures shown in Fig. \ref{fig:Nvortex}a, while the solid line
gives the analytical $\ec$ calculated on the lattice using Eqs.
(\ref{eq:r0}) and (\ref{eq:DeltaF}), multiplied by a constant in order
to account for the unknown proportionality in Eq. (\ref{eq:Ecest}).
We have found an agreement between the analytical and numerical values
for $\ec$ over a wide range of parameters using a proportionality
constant in the range $6.5-8$.

\begin{figure}[t]
  \centering
  \includegraphics[width=\linewidth]{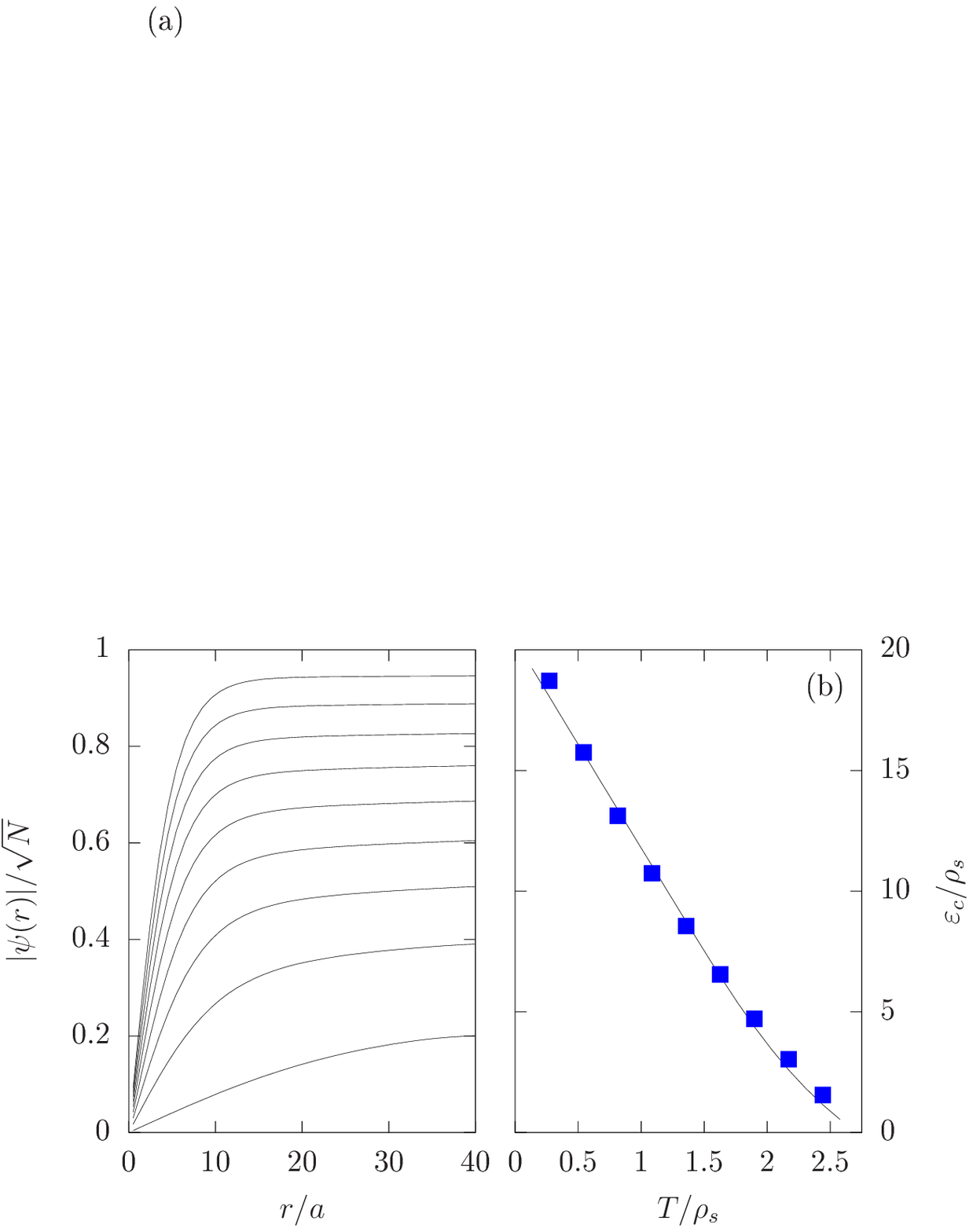}
  \caption{(a) Amplitude of the SC field, $\psi$, as a function of the
    distance from a vortex core, for a number of temperatures below
    $T_{MF}$, obtained by numerically solving Eqs. (\ref{eq:SP}) and
    (\ref{eq:SPpsi}) in polar coordinates. Here we take $N=6$,
    $\lambda=1$ and $ga^2=0.03$, where $a$ is a short distance
    cutoff. (b) Squares: vortex core energies, $\ec$, for the vortices
    depicted in (a). Solid curve: the core energy calculated using
    Eqs. (\ref{eq:r0}) and (\ref{eq:DeltaF}), where a numerical
    proportionality constant was selected, such that the curve agrees
    with the numerical solution.}
  \label{fig:Nvortex}
\end{figure}

\section{Diamagnetism and Nernst effect}
\label{sec:nernst}

Knowing the structure of a vortex, its core energy, and size, it is
now possible to calculate the temperature dependence of the
magnetization, $M_z$, and transverse thermoelectric transport
coefficient, $\alpha_{yx}$, in the phase fluctuations regime. It has
been shown\cite{Nernst-vortex,Halperin,Benfatto-magnetic,Oganesyan},
within a Debye-H\"uckle theory for the thermally excited vortices,
that
\begin{equation}
  \label{eq:mag}
  M_z=-\frac{TB}{\phi_0^2n_f},
\end{equation}
where $\phi_o=h c/ 2e$ is the flux quantum, $B$ the magnetic field,
and $n_f$ is the density of thermally excited vortices. Similarly,
$\alpha_{yx}$, which relates the linear response of an electric
current $J^e_y$ to a transverse thermal gradient $\partial_xT$ via
$j^e_y=\alpha_{yx}(-\partial_xT)$, is given, in the Debye-H\"uckle
regime, by\cite{Nernst-vortex}
\begin{eqnarray}
  \label{eq:alpha}
  % \alpha_{yx}&=&-\frac{2ek_B}{h}\frac{B}{n_f\phi_0}
  % \frac{\ec}{k_B T}
  \alpha_{yx}&=&-\frac{c\ec B}{\phi_0^2Tn_f}
  = \frac{\ec}{T}\frac{cM_z}{T}.
\end{eqnarray}
The Debye-H\"uckle approximation is applicable in the range of
temperatures above, but not to close to $T_{BKT}$, and at the same
time low enough such that the distance between vortices is larger than
their size. In this regime, and in the limit of small magnetic field,
$B\to 0$, the density of thermally excited vortices
is\cite{MinnhagenRMP}
\begin{equation}
  \label{eq:nf}
  n_f\simeq 2r_0^{-2}e^{-2\ec/T}.
\end{equation}
In principle, the $\ec$ that one should use to determine $n_f$ is the
renormalized core energy, which includes also the effect of
fluctuations at short distances below the Debye-H\"uckle screening
length.  However, outside the critical regime close to $T_{BKT}$ the
renormalized $\ec$ is roughly of the same order as its bare value.
Since both $M_z$ and $\alpha_{yx}$ are inversely proportional to
$n_f$, which exhibits an Arrhenius behavior, they both decay strongly
with temperature.  We take such a behavior as a signature of thermally
excited vortices.  It is important to note, though, that these
features are not expected to appear so clearly in simulation because
of finite size effects.

At high temperatures, $T>T_{MF}$, $M_z$ is approximately given by the
$N\to\infty$ limit applied also to the SC fields.\cite{Hayward1} It is
also possible to carry out a similar $N\to\infty$
calculation\cite{nlsm-langevin} of $\alpha_{yx}$. The corresponding,
high temperature results are given by
\begin{equation}
  \label{eq:maght}
  M_z=-\frac{\pi TB}{3\phi_0^2\Delta},
\end{equation}
and
\begin{equation}
  \label{eq:alphaht}
  \alpha_{yx} = -\frac{\pi cB}{6\phi_0^2\Delta} = \frac{cM_z}{2T},
\end{equation}
where $\Delta$ is a solution to the following saddle-point equation
\begin{equation}
  \label{eq:SPht}
  \frac{N-2}{\beta\rhos}\int\frac{d^2p}{(2\pi)^2}\frac{1}{\lambda p^2+g+\Delta}
  +\frac{2}{\beta\rhos}\int\frac{d^2p}{(2\pi)^2}\frac{1}{p^2+\Delta} = N.
\end{equation}
Eqs. (\ref{eq:maght}) and (\ref{eq:alphaht}) are applicable as long as
$\Delta a^2<1$.  In this regime, and for the case $\lambda \geq1$ and
$ga^2<1$, Eq. (\ref{eq:SPht}) gives
\begin{equation}
  \label{eq:Delta}
  \Delta a^2=32\left(\frac{ga^2}{32\lambda}\right)^{\frac{N-2}{N-2+2\lambda}\frac{T_{MF}}{T}}.
\end{equation}
%where the condition $\Delta a^2<1$ implies that this solution is valid in the range
%$1<T/T_{MF}<(N-2)\ln(32\lambda/ga^2)/(N-2+2\lambda)\ln(32)$.

Therefore, we conclude that the rapid decay of both $M_z$ and
$\alpha_{yx}$ in the phase fluctuations regime should crossover to a
much slower decay as the temperature is increased through
$T_{MF}$. Fig. \ref{fig:interp1} demonstrate this point by showing
$-M_z$ and $-\alpha_{yx}$ for a square lattice with $\lambda=1$ and
$ga^2=0.03$. For these parameters $T_{BKT}\approx 1.8\rhos$ and
$T_{MF}\approx 2.7\rhos$.  The phase fluctuations segment is based on
Equations (\ref{eq:mag}) and (\ref{eq:alpha}), while the high
temperature segment is based on Equations (\ref{eq:maght}) and
(\ref{eq:alphaht}). To determine $n_f$ we use in Eq. (\ref{eq:nf}) the
temperature dependent $\ec$ as calculated in the previous section.  We
terminate the phase fluctuations segment when $n_fr_0^2\approx 0.35$,
since at higher temperatures the vortices are no longer distinct
objects, and the Debye-H\"uckle approximation is expected to fail. A
schematic guide to the eye is depicted by the dashed curves, which
interpolate the crossover between the phase fluctuations and Gaussian
fluctuations regimes.

\begin{figure}[t]
  \centering
  \includegraphics[width=\linewidth]{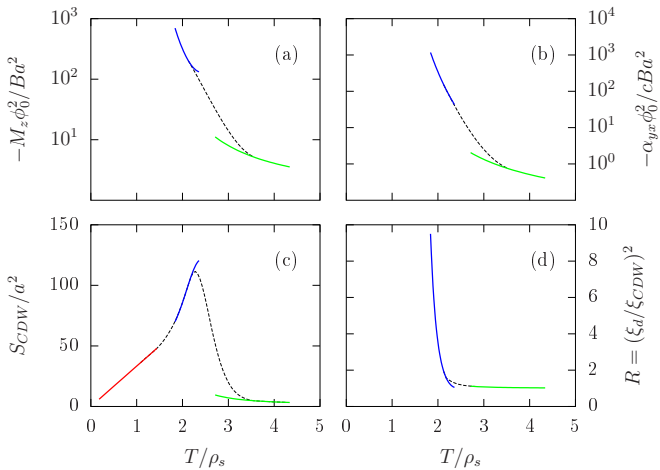}
  \caption{(a) Magnetization, (b) transverse thermoelectric transport
    coefficient, (c) X-ray structure factor, and (d), the unitless
    ratio $R$, all as a function of temperature. Solid lines are based
    on analytic calculations, as explained in the main text, while the
    dashed lines are schematic interpolations in the crossover
    regions, intended as guides to the eye. These results were
    obtained for a square lattice of spacing $a$, $N=6$, $\lambda=1$
    and $ga^2=0.03$.}
  \label{fig:interp1}
\end{figure}

\section{X-ray structure factor}
\label{sec:xray}

Recent X-ray scattering experiments show a pronounced maximum of the
signal at certain incommensurate wave vectors, as a function of
temperature. Hayward \etal\cite{Hayward1} reproduced this maximum
using Monte Carlo simulations of their NLSM. In addition, they were
also able to demonstrate analytically that a maximum exits, by
applying a $1/N$ expansion to their model. Here we use our approach to
show that the signal increases with temperature in a range of
temperatures below $T_{MF}$, and decreases above. To do so, we
calculate the CDW correlation function
\begin{eqnarray}
  \label{eq:Gdef}
  G_{\alpha\beta}(\br,\br') & = & \braket{n_\alpha(\br)n_\beta(\br')}
  \nonumber \\ &  = & \frac{1}{Z}\int\D\psi\D\psi^*
  \D\na\D\bar\sigma\,n_\alpha(\br)n_\beta(\br')\nonumber \\
  &\times&e^{-\beta H +i \int \dr \,\bar\sigma\left(|\psi|^2
      +\sum_\alpha n_\alpha^2-N\right)},
\end{eqnarray}
within our saddle-point approximation, where the integral over
$\bar\sigma$ is replaced by its saddle-point value, given by
Eq. (\ref{eq:SP}), for each configuration of $\psi$.
  
Consider first low temperatures, $T<T_{BKT}$, where essentially there
are no vortices. Ignoring SC amplitude fluctuations, which are
expected to be small, we approximately have $\psi(\br)=\psi_0$, which
gives $\sigma=0$. Hence,
\begin{eqnarray}
  \label{eq:Gp}
  G_{\alpha\beta}(\bp)& = & \int \dr\,e^{-i\bp\cdot\br}G_{\alpha\beta}
  (\br,0) \nonumber \\ & \approx & \frac{\delta_{\alpha\beta}}
  {\beta\rhos(\lambda p^2+g)}.
\end{eqnarray}
The X-ray structure factor, $S_{CDW}$, is obtained by taking the
$\bp\to 0$ limit of $G_{\alpha\beta}(\bp)$ with $\alpha=\beta$, thus,
\begin{equation}
  \label{eq:SCDW}
  S_{CDW}(T<T_{BKT}) \approx \frac{T}{\rhos g},
\end{equation}
which grows linearly with temperature.

At higher temperatures, thermally excited vortices appear in the
system. In principle, one should average over such vortex
configurations, each with its corresponding saddle-point solution,
$\sigma$. Instead, we estimate their effect on $S_{CDW}$ by ignoring
their spatial distribution and considering only their reduction of the
average value of $|\psi|^2$.  Specifically, we replace $|\psi|^2$ in
Eq. (\ref{eq:SP}), by its spatial average,
\begin{equation}
  \label{eq:psiavg}
  \overline{|\psi|^2} \approx |\psi_0|^2-0.45\, n_fr_0^2,
\end{equation}
where the second term accounts for the vanishing $\psi$ inside the
vortex cores. The numerical factor, $0.45$, was extracted from the
numerical solutions of the vortex structure, which were described
above in Section \ref{sec:core}. Since $S_{CDW}$ is a long-wavelength
quantity we take the solution,
\begin{equation} 
\label{eq:sigma0}
\widetilde\sigma=\frac{g\rhos}{2T}\left[\left(\frac{ga^2}{32\lambda}\right)
^{\left(1-0.45n_f r_0^2\right)\left(\frac{T_{MF}}{T}-1\right)}-1\right],
\end{equation}
of the resulting saddle-point equation as an approximation for all
vortex configurations. This results in
\begin{equation}
  \label{eq:SCDWv}
  S_{CDW}%(T_{BKT}<T<T_{MF})
  \approx \frac{T}{\rhos g + T\widetilde\sigma}.
\end{equation}
Since $\widetilde\sigma<0$, we find that $S_{CDW}$ curves upward for
temperatures above $T_{BKT}$ but sufficiently below $T_{MF}$, such
that vortices are distinct objects.

Above $T_{MF}$ the fluctuations become approximately Gaussian, and
their effects can be obtained using the $N\to\infty$ limit also on the
SC order parameter, as is described by Hayward \etal.\cite{Hayward1}
Accordingly, the X-ray structure factor in the high temperature phase
is given by
\begin{equation}
  \label{eq:Scdwht}
  S_{CDW}(T>T_{MF}) \approx \frac{T}{\rhos(g+\Delta)}.
\end{equation}
where, as before, $\Delta$ is the solution of Eq. (\ref{eq:SPht}).
The three solid segments in Figure \ref{fig:interp1}c depict $S_{CDW}$
in the low temperature SC phase, the phase fluctuations regime, and
the high temperature regime, as given respectively by Eqs.
(\ref{eq:SCDW}), (\ref{eq:SCDWv}) and (\ref{eq:Scdwht}).  We therefore
conclude that the maximum in the structure factor may be viewed to
occur at the crossover from a phase fluctuations regime to a high
temperature, Gaussian fluctuations regime.

From the correlation function, Eq. (\ref{eq:Gdef}) it is also possible
to extract the CDW correlation length, $\xi_{CDW}$. When $\sigma$ is
approximately uniform, $G_{\alpha\beta}(\bp)$ has a Lorentzian form,
whose width is defined to be $\xi_{CDW}^{-1}$, \ie,
$G_{\alpha\beta}(\bp)\sim(p^2+\xi_{CDW}^{-2})^{-1}$, or,
$\xi_{CDW}^2=\lambda\rhos S_{CDW}/T$.  Hayward \etal\cite{Hayward2}
considered a dimensionless ratio between this correlation length and
another length scale, $\xi_d$, which can be extracted from the
diamagnetic magnetization by
\begin{equation}
  \label{eq:xid}
  \xi_d^2 = -\frac{3\phi_0^2 M_z}{\pi T B}.
\end{equation}
In the Gaussian limit, this is simply the SC correlation length while
in the fluctuating vortices regime it roughly measures the distance
between vortices. Using Monte Carlo simulations, Hayward \etal~showed
that the dimensionless ratio,
\begin{equation}
  \label{eq:Rdef}
  R(T)=\left(\frac{\xi_d}{\xi_{CDW}}\right)^2
\end{equation}
decreases with temperature. Following our approach, we plot $R(T)$ in
Figure \ref{fig:interp1}d, in the phase fluctuations and high
temperature regimes, with a schematic interpolation between them. At
high temperatures described by Gaussian fluctuations one expects
$R(T\to\infty)=1/\lambda$. The experimental results for $R$ in
Ref. \onlinecite{Hayward2} indicate that $R<1$ in this limit,
therefore possibly implying $\lambda>1$.

\section{Discussion}
\label{sec:conc}

The first question one must address when trying to apply our results
to the $O(6)$ NLSM considered in Ref. \onlinecite{Hayward1}, is
whether the saddle-point approximation, appropriate when $N\to\infty$,
is applicable to a model with a finite number of CDW fields. In the
relevant model, there are four CDW fields, which do not, in two
dimensions, order at any finite temperature. Without any CDW phases
aside from a simple disordered phase, it is reasonable to expect that
the saddle-point approximation captures the behavior of the CDW
fields. Finite-$N$ corrections may, however, alter the numerical
details of the solution, which can introduce discrepancies between the
simulations and our results.  On the other hand, our results would not
be applicable in the presence of long range CDW order, which may, in
principle, occur in layered systems with strong enough coupling
between the layers.

As alluded to in the Introduction, the question of whether the cuprate
high-temperature superconductors actually exhibit significant thermal
fluctuations in the form of vortices, has been under debate. Here, we
would like to ask what are the consequences of making such an
assumption on the parameters that enter the NLSM. Consider first the
range of temperatures above $T_{BKT}$ where one expects to find
signatures of thermally excited vortices. Our results indicate that
this range grows with $\lambda$. Experimentally, the large range of
temperatures above $T_c$ in which there is a strong Nernst signal in a
number of underdoped cuprates, has been advocated\cite{Onglongprb} as
evidence for the existence of thermally excited vortices in these
systems. Quantitatively, it was claimed that phase fluctuations may
exist up to at least $3T_c$ in \LSCO and up to almost $2T_c$ in
\BSCCO. This implies that one would need to take $\lambda>1$ in order
to account for the large phase fluctuations regime.  As we have noted
above, experimental results for $R(T)$ may indicate that $\lambda>1$
in \YBCO~as well. On the other hand, Hayward \etal\cite{Hayward1}
reproduced the X-ray structure factor maximum as a function of
temperature using $\lambda=1$.

An additional point of comparison with experiments is the value of
$\ec$ at temperatures just above $T_{BKT}$. For $\lambda=1$ and
$ga^2=0.3$ we find the ratio $\ec(T \gtrsim T_{BKT})/T_{BKT} \approx
3$, which increases for larger $\lambda$. Analyzing Nernst
measurements from the point of view of vortex fluctuations, we have
estimated\cite{Nernst-vortex} that in \LSCO $\ec/T_c\approx 4-5$,
which is consistent with $\lambda>1$.  Within the context of our
calculation it seems that $\ec/T_c$ cannot be any larger in the
NLSM. However, an analysis\cite{Benfatto-Bilayer} of finite-frequency
sheet conductivity in underdoped
Y$_{1-x}$Ca$_x$Ba$_2$Cu$_3$O$_{7-\delta}$ films has concluded that
$\ec/T_c\approx 8$.  In any event, it seems that $\ec$ is much smaller
than its BCS value, which is of the order of the Fermi energy. By
construction, the NLSM contains this piece of phenomenology, as its
energetics is set by $\rhos$.  However, to understand this fact one
must consider the microscopic details at the basis of the
phenomenological model.

\acknowledgements

This research was supported by the Israel Science Foundation (Grant
No. 585/13).


\begin{thebibliography}{999}
\bibitem{constraint} We choose to normalized the fields to $N$, which
  should be compared with the convention used by Hayward
  \etal\cite{Hayward1}, who normalize the fields to $1$. As a result,
  when comparing our results with their's, one needs to rescale
  $\rhos$ by $N$.
\end{thebibliography}
\end{document}